\documentclass[journal,twoside,web]{ieeecolor}
\usepackage{tmi}
\usepackage{cite}
\usepackage{amsmath,amssymb,amsfonts}
\usepackage{algorithmic}
\usepackage{graphicx}
\usepackage{textcomp}

\usepackage{subcaption}

\usepackage{booktabs}                   
\usepackage{multirow}                   
\usepackage{diagbox}                    
\usepackage{array}                      
\usepackage{longtable}                  
\usepackage{paralist}                   
\usepackage{xltabular}
\usepackage{bbm}
\usepackage{makecell}
\usepackage[utf8]{inputenc}
\usepackage{pifont}

\def\BibTeX{{\rm B\kern-.05em{\sc i\kern-.025em b}\kern-.08em
    T\kern-.1667em\lower.7ex\hbox{E}\kern-.125emX}}

\begin{document}
\title{Whole-brain Transferable Representations from Large-Scale fMRI Data Improve Task-Evoked Brain Activity Decoding}

\author{Yueh-Po Peng, Vincent K.M. Cheung, and Li Su
\thanks{This work was supported in part by the National Science and Technology Council, Taiwan, under grant MOST 110-2221-E-001-010-MY3.}
\thanks{Y.-P. Peng is with Gamania Digital Entertainment Co., Ltd., Taipei 114 Taiwan. This work was conducted during his prior affiliation with the Institute of Information Science, Academia Sinica, Taipei 115 Taiwan (e-mail: yuehpo@iis.sinica.edu.tw).}
\thanks{V.K.M.C is with Sony Computer Science Laboratories, Inc., Tokyo, 141-0022 Japan
(e-mail: cheung@csl.sony.co.jp).}
\thanks{L.S. is with the Institute of Information Science, Academia Sinica, Taipei 115 Taiwan (e-mail: lisu@iis.sinica.edu.tw).}}

\maketitle

\begin{abstract}
A fundamental challenge in neuroscience is to decode mental states from brain activity. While functional magnetic resonance imaging (fMRI) offers a non-invasive approach to capture brain-wide neural dynamics with high spatial precision, decoding from fMRI data---particularly from task-evoked activity---remains challenging due to its high dimensionality, low signal-to-noise ratio, and limited within-subject data. 
Here, we leverage recent advances in computer vision and propose \textit{STDA-SwiFT}, a transformer-based model that learns transferable representations from large-scale fMRI datasets via spatial-temporal divided attention and self-supervised contrastive learning. 
Using pretrained voxel-wise representations from 995 subjects in the Human Connectome Project (HCP), we show that our model substantially improves downstream decoding performance of task-evoked activity across multiple sensory and cognitive domains, even with minimal data preprocessing.
We demonstrate performance gains from larger receptor fields afforded by our memory-efficient attention mechanism, as well as the impact of functional relevance in pretraining data when fine-tuning on small samples.
Our work showcases transfer learning as a viable approach to harness large-scale datasets to overcome challenges in decoding brain activity from fMRI data.
\end{abstract}


\section{Introduction}
\label{sec:introduction}

The goal of brain decoding is to infer mental states from neural activity. Decoding models are not only widely employed in neuroscience research \cite{haynes2015primer}, but also form the backbone of many brain-computer interfaces (BCIs) that improve the well-being of users by overcoming physical limitations, such as communication with locked-in patients \cite{owen2006detecting} or facilitating stroke-recovery \cite{cervera2018brain}. 

One popular approach for brain decoding is to use blood oxygen-level dependent (BOLD) activity recorded from functional magnetic resonance imaging (fMRI) as input features. While fMRI offers excellent (sub-)millimeter spatial resolution, it suffers several limitations: First, 
fMRI data is high-dimensional---often exceeding millions of voxel measurements when considering both time and space, but suffers from low signal-to-noise ratio as only task-related activity changes of $\sim$1-5\% are observed \cite{parrish2000impact}. This presents a serious curse-of-dimensionality problem \cite{bellman1957dynamic} and demands high computational resources. Second, the poor temporal resolution of BOLD responses ($\sim6 s$ between peak and stimulus onset) means that rapid changes in neural activity may be difficult to disentangle. Third, brain decoding models are hard to generalize  across subjects due to individual variability in neural activity and anatomy. This is particularly the case for task fMRI as opposed to resting-state fMRI, as idiosyncratic task differences introduce additional challenges that hinder decoding transferability. Consequently, existing decoding models are typically restricted to classifying mental states from tasks specific to its training data, and rely on voxels selected from \textit{a priori}-defined regions-of-interest (ROI) or aggregated activity from functionally/anatomically homogeneous regions (or brain \textit{parcels}). However, ROI- and parcellation-based decoding entail three critical disadvantages \cite{cheung2023decoding}: First, defining ROIs or parcels requires extensive \textit{a priori} domain knowledge. This is especially problematic if the decoding task is novel and the underlying cognitive processes are not well-defined. Second, brain regions outside of selected regions may contain additional task relevant information for decoding. Third, information aggregation may involve excessive data processing and tradeoff in data granularity. These highlight the need for \textit{whole-brain}-based decoding models.

Recent advances in self-supervised learning (SSL) techniques have demonstrated remarkable performance in
enabling models to learn representations transferable to other datasets from unlabeled data. This opens exciting novel opportunities for representation learning in fMRI data, particularly for end-to-end, whole-brain methods that could overcome issues faced in conventional ROI- and parcellation-based methods. To the best of our knowledge, no simple yet effective system for whole-brain task-fMRI decoding currently exist.

To this end, we propose \textbf{STDA-SWiFT}, a Swin Transformer-based decoding model that effectively learns transferable features from large-scale fMRI datasets to improve performance on downstream decoding tasks. Our model operates on minimally preprocessed whole-brain fMRI images to allow for a fully end-to-end fine-tuning workflow. Inspired from recent advances in video understanding \cite{gberta_2021_ICML, wang2022long}, we also introduce a space-time divided attention (STDA) mechanism for the Swin Transformer architecture. This memory-efficient configuration separately models spatial and temporal dimensions to reduce computational overhead while preservING functional locality. Compared to joint-spatiotemporal (4D) attention designs, our architecture better matches the structure of fMRI data and allows for larger spatial window sizes under limited computational resources. 

Furthermore, under a SimCLR-based contrastive learning framework \cite{chen2020simple}, we systematically investigate the impact of different data augmentation strategies on pretraining and fine-tuning performance. While SSL techniques highly rely on data augmentation, data augmentation strategies for fMRI data is not well investigated. In fact, fMRI data differs significantly from natural images or videos: it is neither location-invariant nor scale-invariant, and exhibits complex spatiotemporal dependencies that are not well-captured by conventional augmentation strategies or architectures. We show which augmentation strategies are beneficial specifically for fMRI data.

\section{Previous Work}

Here, we briefly review existing approaches to decode task-evoked brain activity from fMRI data.

\subsection{ROI-based decoding}
Region of Interest (ROI)-based methods are among the earliest and most widely adopted approaches in task-based fMRI.
These methods involve selecting specific anatomical or functional brain regions based on prior neuroscientific knowledge or independent localizer tasks, and analyzing neural activity within those constrained areas. By isolating brain regions known to support particular cognitive functions, ROI-based methods enable hypothesis-driven investigations with high interpretability, as well as provide a simple method for dimension reduction when building decoding models \cite{haynes2006decoding}.

Classic studies such as Haxby \emph{et al.} \cite{haxby2001distributed} laid the foundation for ROI-based multivoxel pattern analysis (MVPA), showing that patterns of activity within the ventral temporal cortex can discriminate between object categories such as faces and houses, even when the mean activation levels do not differ. Likewise, Haynes and Rees 2005 \cite{haynes2005predicting}, and Kamitani and Tong 2005 \cite{kamitani2005decoding} demonstrated orientation biases in neurons in the early visual cortex below the conventional spatial resolution of fMRI by exploiting the spatial patterns of neighboring voxels. The searchlight method generalizes ROI-based decoding to the whole-brain by iteratively training a decoder using a small local neighborhood across all voxels \cite{haynes2015primer}.

Several brain computer interfaces also exploit task-related changes in target ROIs as their method of control. For example, a matrix speller by Sorger \emph{et al.} \cite{sorger2012real} encodes letters from ROIs activated by motor imagery, mental calculation, or inner speech. Similarly, differential activity in the supplementary motor area when imagining playing tennis versus navigating at home has been used to demonstrate intention and communication in a vegetative-state patient \cite{owen2006detecting}. Furthermore, decoded fMRI neurofeedback (DecNef), a technique where participants learn to implicitly up- or down-regulate activity of a particular ROI \cite{shibata2019toward}, has been used for fear reduction training \cite{koizumi2016fear} and modulating facial preference \cite{shibata2016differential}.

In the context of large-scale datasets such as the Human Connectome Project (HCP) \cite{barch2013function}, ROI-based analyses have been extensively applied to decode brain responses across a wide variety of tasks, including working memory, language processing, motor execution, and social cognition. For example, Barch \emph{et al.} (2013) \cite{barch2013function} used task-based fMRI data from HCP to link activation patterns in regions such as the dorsolateral prefrontal cortex (dlPFC) and temporoparietal junction (TPJ) to behavioral performance in cognitive control and social tasks. Similarly, Tavor \emph{et al.} (2016) \cite{tavor2016task} leveraged ROI-defined connectivity patterns during rest to predict individual differences in task activation, which demonstrated the predictive utility of ROI features even in task-free paradigms.

Apart from classification and regression, features from predefined ROIs have also been used for generative tasks. For example, several studies have decoded activity from the visual cortex to reconstruct seen images \cite{shen2019deep, takagi2023high}, as well as the auditory cortex to presented sounds \cite{denk2023brain2music, park2025natural} using deep neural networks. 

\begin{figure*}[t]
    \centering
    \begin{subfigure}{0.43\textwidth}
        \raggedleft
        \includegraphics[width=\linewidth]{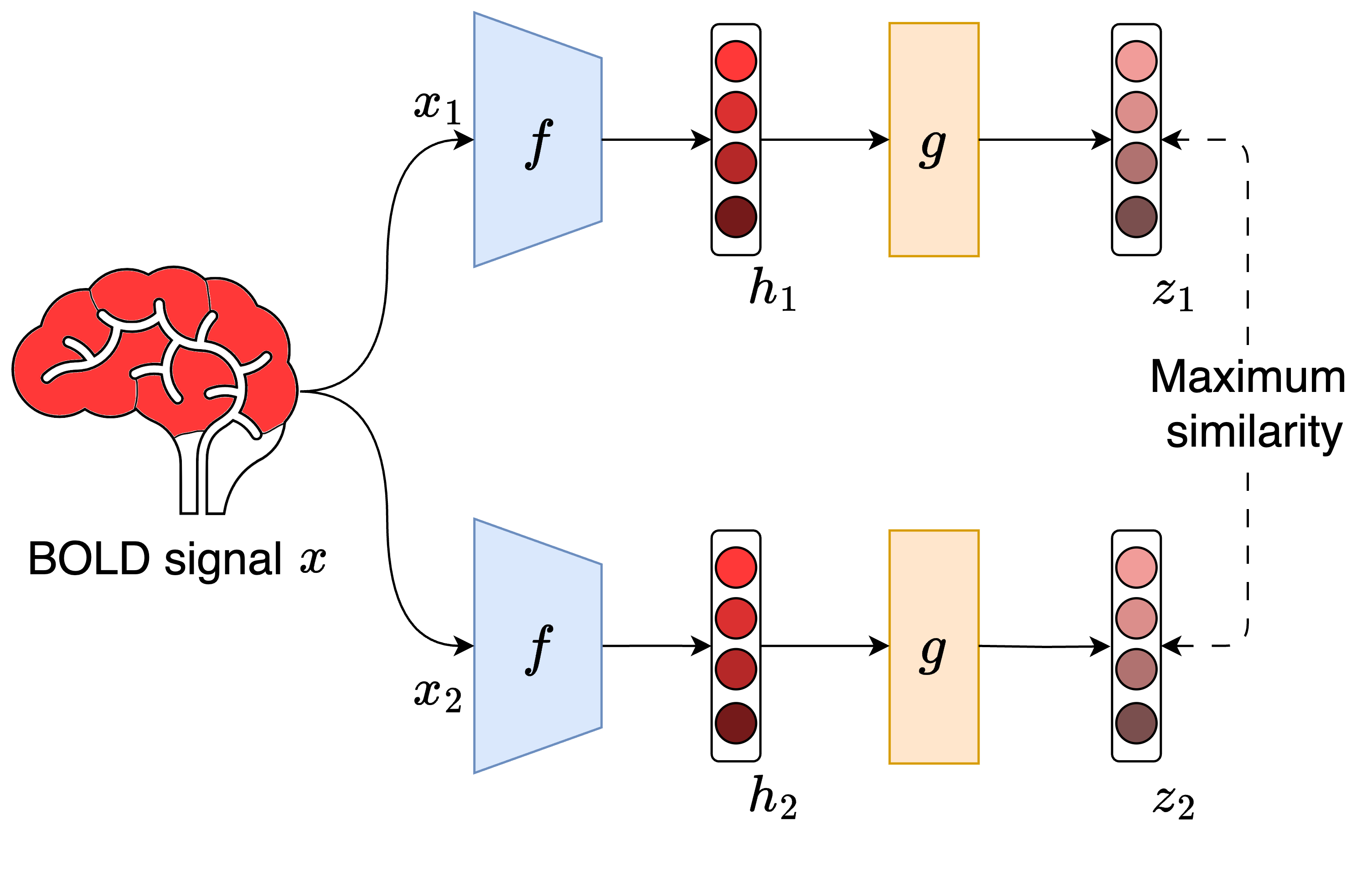}
        \caption{Contrastive learning pipeline for fMRI.}
        \label{fig:contrastive_learning}
    \end{subfigure}
    \hfill
    \begin{subfigure}{0.56\textwidth}
        \raggedright
        \includegraphics[width=\linewidth]{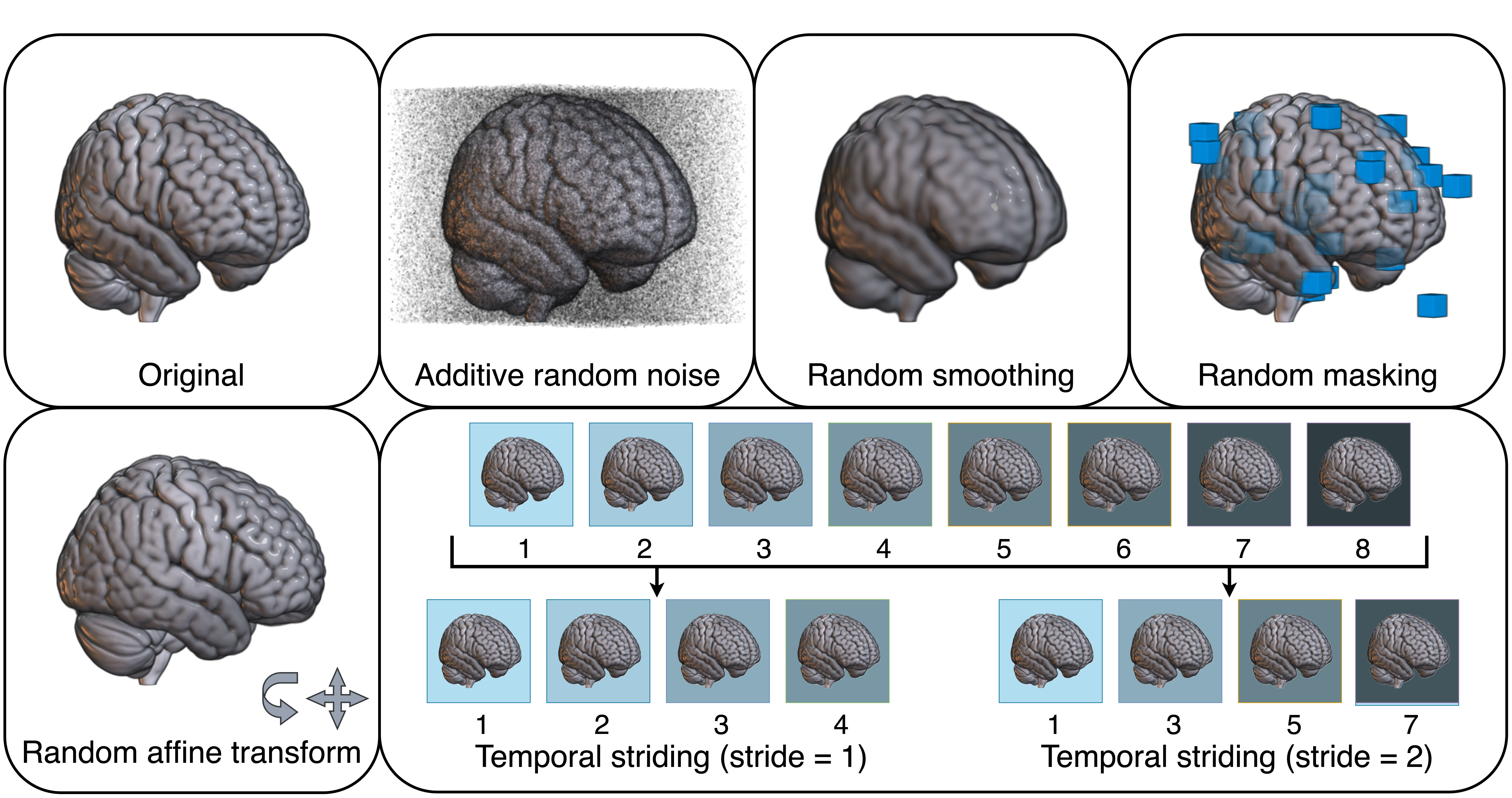}
        \caption{fMRI data augmentation strategies.}
        \label{fig:fMRI_aug}
    \end{subfigure}
    \caption{Illustration of (a) the contrastive learning framework used to pretrain encoders using augmented BOLD signal pairs, and (b) augmentation strategies applied to fMRI.}
    \label{fig:combined_teaser}
\end{figure*}

\subsection{Parcellation-based decoding}
Parcellation-based methods segment the whole-brain into functionally or anatomically homogeneous regions---or parcels---for decoding. This approach dramatically reduces the number of decoding features from $\sim10^5$ voxels to $\sim10^2$ parcels. Notable parcellations examples (see \cite{lawrence2021standardizing} for a curated list) include Yeo \cite{yeo2011organization} that is based on functional coupling between regions from resting-state fMRI data, Glasser \cite{glasser2016multi} that subdivides the brain into 360 regions based on function, connectivity, and anatomy, Harvard-Oxford \cite{rushmore2022hoa2} that parcellates the neocortex into 48 regions along its principle gyri, and Schaefer \cite{schaefer2018local} that is derived from gradient-weighted Markov random fields from resting-state and task fMRI data. 

Two prominent studies take a parcellation-based approach to decoding from task fMRI data. In Thomas \emph{et al.} \cite{thomas2022self}, the brain is parcellated into a sparse dictionary matrix of functional networks using a scheme known as Dictionaries of Functional Modes (DiFuMo). While this enabled model pretraining on task-evoked fMRI data in the HCP dataset using SSL techniques inspired from natural language processing (NLP), the parcellation scheme required extensive preprocessing . 
On the other hand, Ortega Caro \emph{et al.} \cite{ortega2023brainlm} introduced a transformer-based fMRI foundation model---the Brain Language Model (BrainLM)---that was trained on 6700 hours of fMRI data parcellated into 424 brain regions from the AAL atlas. Utilizing self-supervised masked-prediction training, BrainLM demonstrated proficiency in both fine-tuning for predicting clinical variables and task-evoked activity from resting-state fMRI. 

\subsection{Whole-brain decoding}
Several recent approaches have emerged to decode mental states by directly feeding the entire fMRI volume into advanced machine learning models. This whole-brain approach harnesses the comprehensive spatial and temporal information encoded across all voxels, and has the benefit of not requiring \textit{a priori} domain knowledge in selecting ROIs or parcels based on task relevance, as well as reduced information loss and preprocessing required when aggregating from multiple voxels. 

One example is an end-to-end whole-brain decoding model by Shi \emph{et al.} \cite{shi2023self}. To train this SSL model, the fMRI sequence recorded during each stimulus presentation is divided into three temporal sections, namely beginning, middle, and end. The model is then trained to differentiate between neighboring and distant temporal segments via contrastive learning. Pretraining on five tasks from the HCP dataset, the model was able to effectively generalize across subjects and showed similar decoding performance in downstream decoding tasks with 12 subjects compared to 100 subjects in a randomly initialized model. 

Another advancement in whole-brain modeling is the development of SwiFT (Swin 4D fMRI Transformer), proposed by Kim \emph{et al.} \cite{kim2024swift}. SWiFT is an adaptation of the Swin Transformer architecture \cite{liu2021swin} for resting-state fMRI data by employing 4D window multi-head self-attention and absolute positional embeddings. Evaluating on large-scale datasets such as HCP and the UK Biobank (UKB) \cite{sudlow2015uk}, SwiFT outperformed other state-of-the-art models in predicting subject phenotype and cognitive traits, as well as task-related brain activity based on resting-state fMRI data \cite{kwon2025predicting}. Nevertheless, despite learning transferable representations for downstream decoding tasks, this 4D Transfomer architecture is extremely memory intensive: an attention field with a 4-voxel edge has size $4^4=256$, and a $6 \times 6 \times 6 \times 6$ attention is already impractical on an Nvidia V100 GPU without significantly reducing training batch size.

Alternatively, rather than using all voxels in the whole brain for decoding, other methods seek to identify task-relevant voxels from whole-brain fMRI data. For example, Cheung \emph{et al.} \cite{cheung2023decoding} proposed a whole-brain feature selection framework based on cross-validation and feature importance using Shapley additive explanations (SHAP) \cite{lundberg2017unified}. Their method identified voxels in the somatosensory cortex that were relevant for decoding pitch in addition to relying on voxels in the auditory cortex.

\section{Method}

In this section, we describe the training approach and architecture of our proposed STDA-SwiFT model.

\subsection{Self-supervised pretraining}

Figure~\ref{fig:combined_teaser} illustrates the overall SSL-based pretraining scheme (Figure \ref{fig:contrastive_learning}) and data augmentation strategies (Figure \ref{fig:fMRI_aug}) used to learn robust representations from whole-brain fMRI. Here, we used the contrastive learning method SimCLR \cite{chen2020simple} for pretraining: 

Let \(B\) be a batch of fMRI images $\{x_i\}^B_{i=1}$ sampled during stimulus presentation. 
Each $x_i$ is of dimension $T\times H\times W\times D\times 1$, where $T$, $H$, $W$, and $D$ represent time, height, width, and depth, respectively. The channel dimension (i.e., the last dimension of $x$) is 1 for the input. Following the standard setting of SimCLR, each sample \(x_i\) is augmented into two views $\{x'_{2i-1}, x'_{2i}\}$, then 
encoded by the encoder \(f(\cdot)\) to obtain the representations denoted as $h_{2i-1}:= f(x'_{2i-1})$ and $h_{2i}:= f(x'_{2i})$.
A projector network \(g(\cdot)\) projects $h_{2i-1}$ and $h_{2i}$ into \(z_{2i-1}:=g(h_{2i-1})\) and \(z_{2i}:=g(h_{2i})\), respectively. The NT-Xent loss \cite{chen2020simple} for the $i$th pair \((z_{2i-1}, z_{2i})\) is defined as:
\begin{equation}
\ell_{2i-1, 2i} := -\log \frac{\exp(\text{sim}(z_{2i-1}, z_{2i}) / \tau)}{\sum_{j=1}^{B} \mathbbm{1}_{j \neq i} \exp(\text{sim}(z_{2i-1}, z_{2j}) / \tau)},
\end{equation}
where \(\text{sim}(\cdot, \cdot)\) is the cosine similarity between two vectors, \(\tau\) is a temperature (where $\tau=0.1$ throughout this paper) 
, and \(\mathbbm{1}_{j \neq i}\) is an indicator function equal to 1 if \(j \neq i\) and 0 otherwise. The NT-Xent loss for the entire batch is then 
\begin{equation}
\mathcal{L}_{\text{batch}} := \frac{1}{2B} \sum_{i=1}^{B} \left( \ell_{2i-1, 2i} + \ell_{2i, 2i-1} \right),
\end{equation}
where the loss is averaged over all positive pairs in the batch, ensuring that each sample contributes equally.

\begin{figure}[t]
    \centering
    \includegraphics[width=\linewidth]{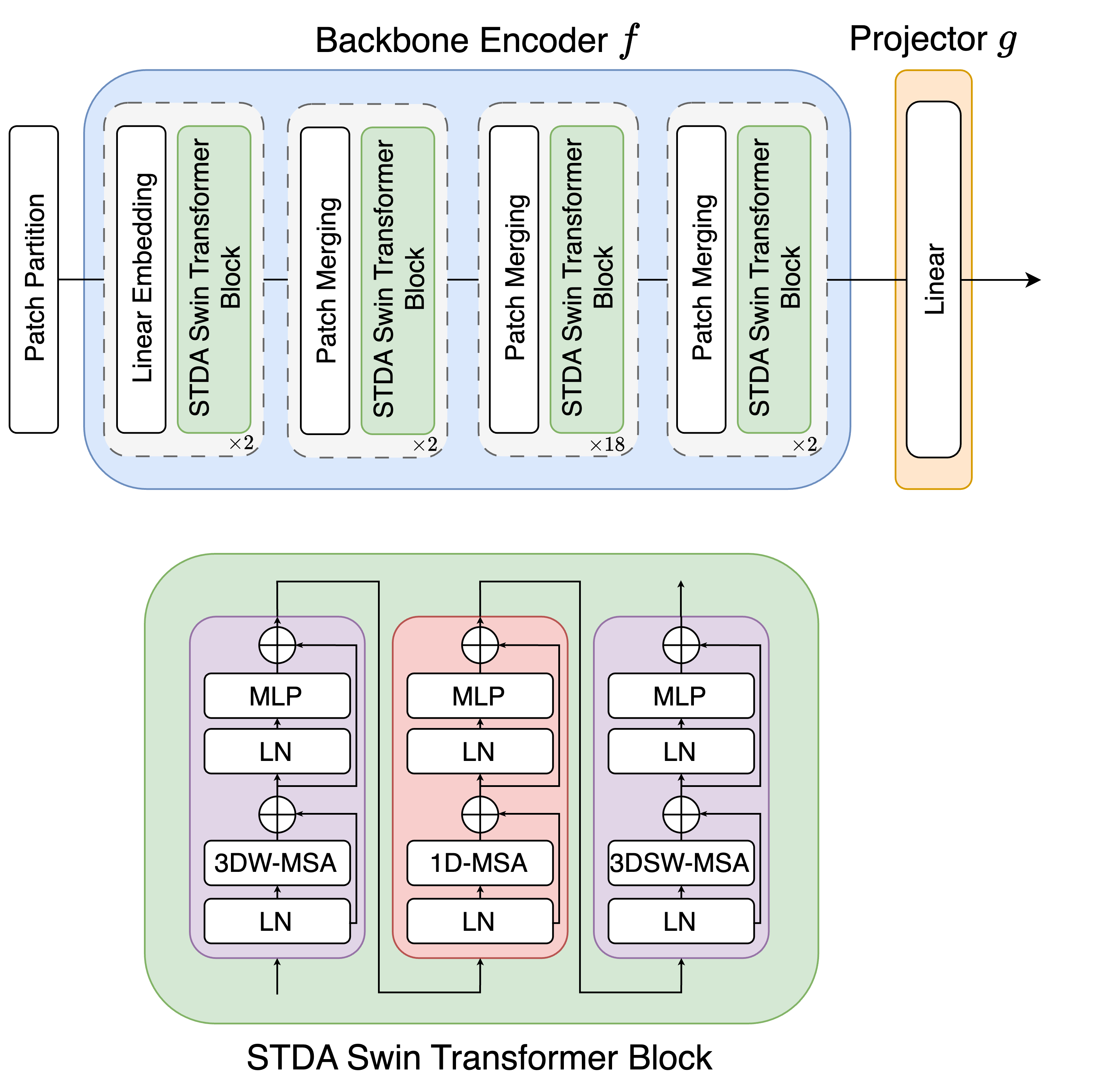}
    \captionsetup{justification=centering} 
    \caption{Swin Transformer architecture with spatio-temporal decoupled attention (STDA) mechanism.}
    \label{fig:model_arch}
\end{figure}

\begin{figure}[t]
    \centering
    \includegraphics[width=\linewidth]{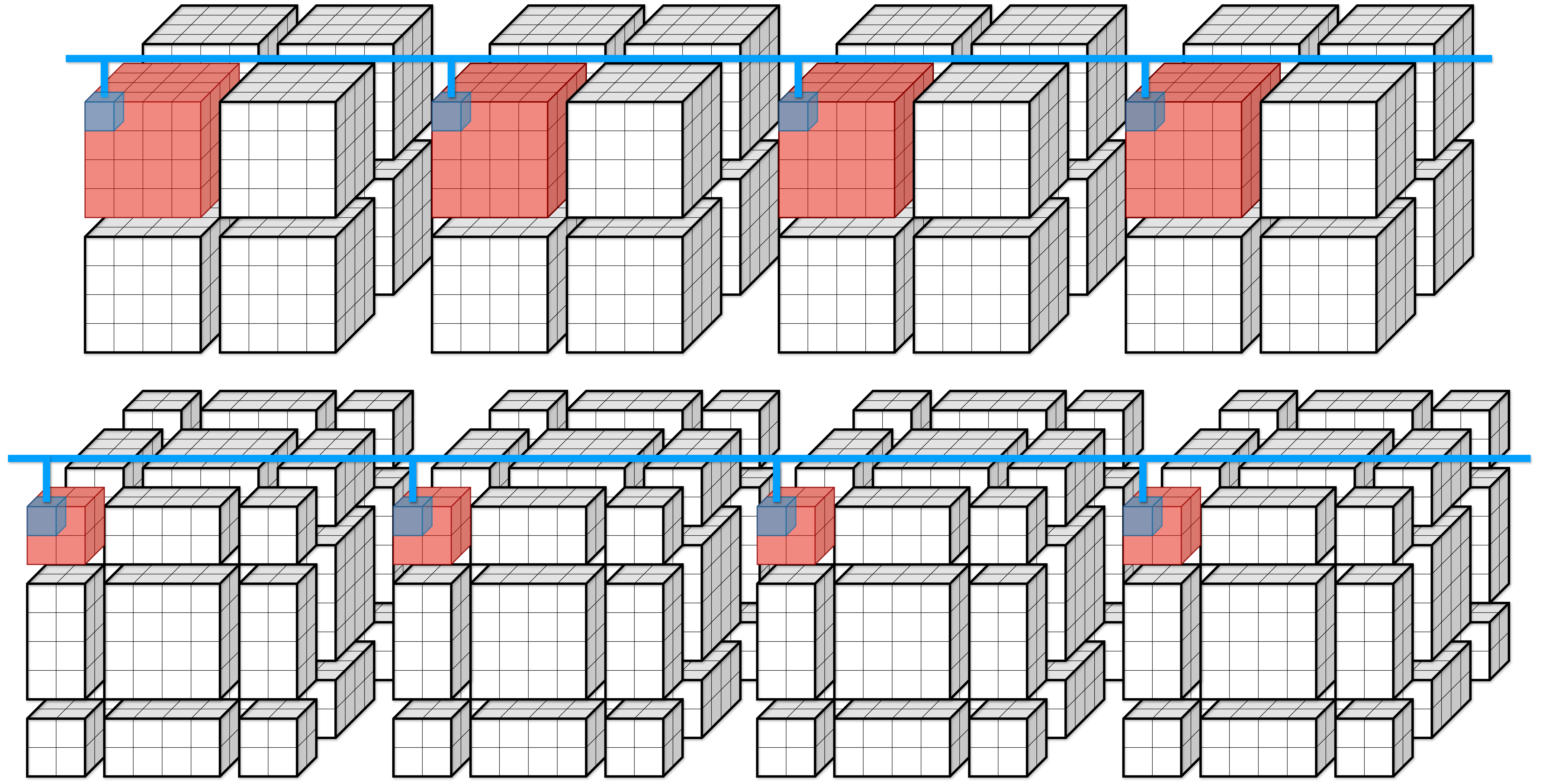}
    \caption{Spatio-temporal decoupled attention (STDA) mechanism for Swin Transformer block.}
    \label{fig:STDA_mechanism}
\end{figure}

\subsection{Model}

Our proposed encoder model $f(\cdot)$ (Figure \ref{fig:model_arch}) is based on the Swin Transformer architecture \cite{liu2021swin}. Unlike SwiFT \cite{kim2024swift} that uses 4D joint space-time attention, we utilize spatio-temporal decoupled attention (STDA) \cite{gberta_2021_ICML} to significantly optimize memory consumption and computation efficiency. For 4D self-attention on a window of size $d\times d\times d\times d$, the computational complexity scales to $O(d^8)$. 
On the other hand, our spatial
attention of size $d\times d\times d$ and the temporal attention of $t$, scales the complexity down to $O(d^6+t)$. A spatial self-attention window of size $6\times 6\times 6$ (length 216) thus occupies less memory in comparison to a 4D self-attention window of size $4\times 4\times 4\times 4$ (length 256).
Apart from memory advantages, another motivation for separating the signal into three spatial dimensions and one temporal dimension is to capitalize on the functional organization of the human brain, where neighboring brain regions perform similar functions and are anatomically similar. Disentangling the spatial dimension into lower dimensions will likely corrupt this characteristic.

Now we describe the architecture of our proposed model in detail. First, each $x_i$ 
is partitioned into non-overlapping patches. The dimension of each patch is $P\times P\times P$ (voxels), such that the total number of the patches for $x_i$ is $THWD/P^3$. 
Each of the patches is projected onto a $C$-dimensional embedding through a linear embedding layer. We set $H=W=D=96$, $P=6$, $C=36$, and $T$ is dataset dependent (see Section \ref{sec:data} for details). Then, a patch merging layer is applied in the spatial dimensions. This has the effect of reducing the spatial dimension by half, while doubling the channel dimension \cite{kim2024swift}. Following the Swin Transformer, we henceforth denote the dimension of a feature tensor as $T\times H'\times W'\times D'\times C'$.

Next, the spatial dimensions (i.e., $H'\times W'\times D'$) of the embedding space are partitioned by non-overlapping windows with size $M\times M\times M$, which results in $T\cdot\lceil\frac{H'}{M}\rceil \cdot \lceil\frac{W'}{M}\rceil \cdot \lceil\frac{D'}{M}\rceil$ windows in total. For each of the two consecutive layers, the window partition is shifted by $(\lfloor\frac{M}{2}\rfloor,\lfloor\frac{M}{2}\rfloor,\lfloor\frac{M}{2}\rfloor)$ from that of the preceding layer. 
The proposed STDA Swin Transformer block (see Figure \ref{fig:STDA_mechanism}) incorporates two separated multi-head self-attention (MSA) mechanisms. 
First, spatial attention operates on each spatial window. Second, temporal attention operates on each patch over different time steps. Representing the $(l-1)$th-layered embedding as $z^{l-1}$, two consecutive Spatial-Temporal Swin Transformer blocks are computed as 

\begin{align}
    z^l_{\text{3D}} &= \text{3D-W-MSA}(\text{LN}(z^{l-1})) + z^{l-1}\,, \\
    \hat{z}^l_{\text{3D}} &= \text{MLP}(\text{LN}(z^l_{\text{3D}})) + z^l_{\text{3D}}\,, \\
    z^l_{1D} &= \text{1D-MSA}(\text{LN}(\hat{z}^l_{\text{3D}})) + \hat{z}^l_{\text{3D}}\,, \\
    z^l &= \text{MLP}(\text{LN}(z^l_{\text{1D}})) + z^l_{\text{1D}}\,, \\
    z^{l+1}_{\text{3D}} &= \text{3D-SW-MSA}(\text{LN}(z^{l})) + z^{l}\,, \\
    \hat{z}^{l+1}_{\text{3D}} &= \text{MLP}(\text{LN}(z^{l+1}_{\text{3D}})) + z^{l+1}_{\text{3D}}\,, \\
    z^{l+1}_{\text{1D}} &= \text{1D-MSA}(LN(\hat{z}^{l+1}_{\text{3D}})) + \hat{z}^{l+1}_{\text{3D}}\,, \\
    z^{l+1} &= \text{MLP}(\text{LN}(z^{l+1}_{\text{1D}})) + z^{l+1}_{\text{1D}}\,, 
\end{align}
where 3D denotes MSA in the spatial domain, 1D denotes MSA in the temporal domain, LN denotes layer normalization, MLP denotes multi-layer perceptron (here, we use two fully-connected layers), and W-MSA and SW-MSA represent the regular windowed MSA and shifted windowed MSA in the Swin Transformer architecture, respectively. 

Now, consider a transposed (embedding dimension first) windowed embedding tensor $z'$ with dimensions $C'\times T\times M\times M\times M$. For $\tilde{z}'$, a spatially flattened embedding of $z'$ with dimensions $C'\times T\times M^3$, the spatial self-attention of one head $a$ at time $t$ is 
\begin{equation}
    s_{t}= \text{softmax}\left(\frac{(W^a_Q \tilde{z}'_{t,:,:})^T W^a_K \tilde{z}'_{t,:,:}}{\sqrt{D_h}}\right)W^a_K \tilde{z}'_{t,:,:}
\end{equation}
for $t\in[1,T]$, trainable parameters $W^a_Q$, $W^a_K$ and $W^a_V$, and feature dimension $D_h$. Similarly, for temporal attention, for each spatial position $h,w,d\in[1,M]$, we have
\begin{equation}
    s_{h,w,d}= \text{softmax}\left(\frac{(W^a_Q z'_{:,h,w,d,:})^T W^a_Kz'_{:,h,w,d,:}}{\sqrt{D_h}}\right)W^a_Vz'_{:,h,w,d,:}
\end{equation}
where the attention output is then the concatenation of all attention heads' $s_t$ for all $t$ (for spatial attention) and $s_{h,w,d}$ for all $h,w,d$. Following \cite{kim2024swift}, we set the number of attention heads to 3, 6, 12, and 24 for the four stages, respectively. 
The output of $f(\cdot)$ is the mean over all patches from 4D to 1D (i.e., the dimension of a batch is reduced from $B\times C\times T\times H\times W\times D$ to $B\times C\times 1$ after averaging). This $C$-dimensional feature is then used for classification. 
Finally, the projector \( g(\cdot) \) is simply a dense layer with input dimension 
\(d_e = 288\) (the output dimension of the SwiFT encoder) and output dimension dimension 
\(d_p = 128\) (the dimension of the projection space). 
\subsection{Data augmentation}
\label{data_aug_hcp}
We considered five data augmentation strategies to improve model training (refer to Figure \ref{fig:fMRI_aug}): 

\begin{enumerate}
\item \textbf{Affine transform} (A). 
    To preserve the inherent characteristics of fMRI data, 
    relatively light parameters were considered: 
    scaling within 10\% of its original size (i.e., scaling between 0.9 and 1.1) and rotation within 10 degrees on each axis (e.g., between -10 and +10 degrees).
    
\item \textbf{Additive Gaussian noise} (N). 
Two levels of noise perturbation were considered. For the \emph{low} setting, the standard deviation (\( \sigma \)) of the random Gaussian noise was sampled from $\mathcal{U}\left(0, 0.1\right)$, a uniform distribution between \( 0 \) and \( 0.1 \). As for the \emph{high} perturbation setting, the $\sigma$ value was sampled from $\mathcal{U}\left(0, 0.5\right)$. 
    
\item \textbf{Smoothing} (S). We applied Gaussian kernel smoothing filters to reduce high-frequency noise and anatomical differences. 
Similar to noise augmentation, we considered two levels of smoothing: For the \emph{low} setting, the $\sigma$ value for the Gaussian kernel was sampled from $\mathcal{U}\left(0, 0.5\right)$, while for \emph{high}, the $\sigma$ value of the kernel was sampled from $\mathcal{U}\left(0, 2\right)$.
    
\item \textbf{Masking} (M). 
We applied random masking to \(20\%\) of the fMRI volumes using a fixed spatial mask of size 4$\times$4$\times$4$\times T$, where \(4\times4\times4\) represents a cubic region in the 3D spatial domain, and \(T\) denotes the number of time frames (such that all time points were masked for the chosen voxels).
    
\item \textbf{Temporal striding} (T).  
    Given a sequence $\{x_i\}_{i=1}^N$ of consecutive fMRI volumes, temporal striding involves selecting a subsequence $\{x_{i_j}\}$, such that $i_{j+1} - i_{j}$ is sampled uniformly from $\{1,2,3\}$. Here, we let $j=\{1,2,3,4,5\}$ and note that $x_{i_1}$ need not equal to $x_1$. Also, note that during training, the positive pairs $\{x_{i_j}\}_{j=1}^5$ and $\{x_{i_k}\}_{k=1}^5$ derived from the same fMRI sequence need not be identical. This strategy enforces the model to learn patterns that occur over different time spans, making it more versatile in analyzing brain activity. 

\end{enumerate}

\section{Experiments}

\subsection{Data}
\label{sec:data}

Two datasets were used in our experiments. The first is the Human Connectome Project (HCP) 3T task-fMRI dataset, a large-scale dataset for studying task-evoked brain activity \cite{barch2013function}. It includes fMRI scans from 995 subjects who completed all seven tasks: working memory, gambling, motor, language, social, relational, and emotion (see \cite{barch2013function} for details). We randomly selected 900 subjects for self-supervised pretraining (HCP pretraining set) and held out 95 for fine-tuning evaluation (HCP held-out set).

The second is the Multi-Domain Task Battery (MDTB) dataset, comprising fMRI scans from 24 participants across 47 task conditions covering cognitive, motor, and affective domains \cite{king2019functional}. Following Thomas \emph{et al.} \cite{thomas2022self}, we grouped related task conditions into 26 mental states.\footnote{These tasks, as labeled by the original authors of the dataset, include: CPRO, GoNoGo, ToM, actionObservation, affective, arithmetic, checkerBoard, emotionProcess, emotional, intervalTiming, landscapeMovie, mentalRotation, motorImagery, motorSequence, nBack, nBackPic, natureMovie, prediction, rest, respAlt, romanceMovie, spatialMap, spatialNavigation, stroop, verbGeneration, and visualSearch \cite{king2019functional}.} Due to data quality, we used preprocessed scans from 23 participants, split into training (11), validation (3), and test (9) sets. We repeated experiments over three random splits.

For preprocessing, we randomly sampled $T = 15$ scans per stimulus from the HCP dataset (regardless of stimulus duration) and retained all $T = 30$ scans per stimulus in MDTB. All scans were clipped to 96$\times$96$\times$96 voxel cubes to standardize input across experiments.

\subsection{Experimental settings}

We set up three experiments to investigate decoding performance via transfer learning from large-scale whole-brain fMRI data.

\subsubsection{Task 1: Comparison of data augmentation strategies}
\label{ssl-pretraining}
A key goal of this paper is to investigate the data augmentation strategies and find their combinations most suitable for SSL-based pretraining and fine-tuning of fMRI data. 
Two scenarios were considered: 1) model pretraining on HCP pretraining set and 2) fine-tuning the pretrained model in 1) on the MDTB dataset. 
The compared data augmentation strategies are listed in Table \ref{table:SSL_aug} (for pretraining on the HCP pretraining set) and Table \ref{tab:mdtb_augmentation_results} (for fine-tuning on the MDTB dataset). 

For pretraining experiments, we trained the proposed STDA-SwiFT with a window size of $6\times6\times6$. We report the highest validation accuracy on the HCP pretraining set over 50 training epochs. Here, we utilized 720 subjects for training, while the remaining 180 subjects were used for validation. Accuracy was computed using a $k$-Nearest Neighbor (kNN) classifier with $k=1$.

Each model was trained for 50 epochs with a learning rate of 0.001 and a batch size of 6. 

For fine-tuning experiments, we continued training the pretrained STDA-SwiFT with window size of $6\times6\times6$ for 20 additional epochs with a learning rate of 0.00005, and a batch size of 12.
We report the classification accuracy for different combinations of augmentation strategies applied to the proposed model, pretrained on the HCP dataset using 995 subjects across 7 tasks. To identify the optimal augmentation strategy, the model was fine-tuned on one subject randomly selected from the 11-subject training set of the MDTB dataset. 
Then, the model was validated on the 3-subject validation and tested on the 9-subject test set (see Section \ref{sec:data} on MDTB dataset). 

\subsubsection{Task 2: Brain decoding with pretrained models}
In this experiment, we investigated the effectiveness of our model pretrained on the HCP pretrained set for decoding on unseen tasks. We considered three classification tasks evaluated on the HCP held-out set:

\begin{itemize}
    \item \textbf{Motor}: Classify five movement types—left/right finger (lh/rh), left/right toe (lt/rt), and tongue (t).
    \item \textbf{Relational}: Binary classification distinguishing between a relational condition (comparing shape/texture relations) versus a control condition (simple attribute matching).
    \item \textbf{Motor vs. Relational}: Binary classification task identifying whether the fMRI data captured brain activity during the motor or relational task.
\end{itemize}

We compared the performance of our proposed model with the following baseline models:

\begin{enumerate}
    \item A ResNet-based model with temporal convolution proposed by Shi~\emph{et al.}~\cite{shi2023self}.
    
    \item SwiFT-small (rfMRI-pretrained, W4): a 4D Swin Transformer pretrained on resting-state fMRI (rfMRI) with weights from~\cite{kim2024swift}, using a $4\times4\times4$ window.
    
    \item SwiFT (R, W6): a larger SwiFT model randomly initialized and trained from scratch, following Swin Transformer~\cite{liu2021swin} scaling rules. We increased the number of attention layers from 12 to 24 (by setting 18 layers in the third block) and used a larger $6\times6\times6$ window.
    
    \item SwiFT (T+M+N+S, W6): same as (3), but pretrained on task fMRI data using temporal striding (T), masking (M), noise (N), and smoothing (S) augmentations.
\end{enumerate}

For our proposed STDA-SwiFT model, we compared the following four variants to examine the effects of window size, pretraining strategy, and data augmentation:

\begin{enumerate}
    \item The STDA-SwiFT model with random initialization and a 4$\times$4$\times$4 window, denoted as ``STDA-SwiFT (R, W4).'' 
    \item The STDA-SwiFT model with random initialization and a 6$\times$6$\times$6 window, denoted as ``STDA-SwiFT (R, W6).'' 
    \item The STDA-SwiFT model pretrained on the HCP pretraining set with temporal striding and random masking, denoted as ``STDA-SwiFT (T+M, W6).'' 
    \item The STDA-SwiFT model pretrained with temporal striding, random masking, additive random noise, and random smoothing, then fine-tuned using a 6$\times$6$\times$6 window size. We refer to this model as STDA-SwiFT (T+M+N+S, W6). This setting represents our full pretraining approach, allowing us to quantify the contributions of each self-supervised strategy. 
\end{enumerate}

Following \cite{shi2023self}, fine-tuning was performed with two training set sizes—12 and 76 held-out subjects—while keeping the validation and test sets fixed at 9 and 10 subjects, respectively. This setup (12/9/10 and 76/9/10 splits) enabled us to evaluate our method's robustness under different levels of data availability.
We fine-tuned the models for 15 epochs with learning rate of 0.0001, a batch size of 12. 
\subsubsection{Task 3: Cross-dataset brain decoding}

In this experiment, we investigated the decoding performance of the 26 mental states of the MDTB dataset using our proposed model and different pretraining strategies. The STDA-SwiFT model pretrained with T+M+N+S augmentations and fine-tuned with M+A+N+S was used throughout this experiment. We fine-tuned the models for 20 epochs with learning rate of 0.00005, a batch size of 6. 

Here, we report classification accuracy and F1 scores, and compared our model with the following models:

\begin{itemize}
    \item ROI-based: model proposed by Shi \emph{et al.} \cite{shi2023self}, which processes signals within a bounding box covering the visual cortex.
    \item Parcellacion-based: NLP-inspired parcellation-based decoding models by Thomas \emph{et al.} \cite{thomas2022self}, including a recurrent encoder-decoder model based on LSTM (denoted as Autocoding), a transformer decoder for Causal Sequence Modeling (CSM), Sequence-BERT, and Network-BERT.      
    \item Whole-brain: family of SwiFT models (see Task 2).
\end{itemize}

In addition, we carried out three ablation studies:

\begin{itemize}
    \item \textbf{Pretraining data diversity}: We compared models pretrained on all seven HCP tasks (denoted as ALL7), as well as on individual tasks (Emotion, Gambling, Language, Motor, Relational, Social, and Working Memory), and models trained from scratch (i.e., no pretraining).
    
    \item \textbf{Fine-tuning dataset size}: Following \cite{thomas2022self}, we evaluated performance using 1, 3, 6, and 11 MDTB subjects for fine-tuning.
    
    \item \textbf{Window size}: We trained STDA-SwiFT using window sizes of $2\times2\times2$, $4\times4\times4$, and $6\times6\times6$, and compared them against SwiFT (trained with a $4\times4\times4$ window) to assess the effect of spatial context. The $2\times2\times2$ setting provides highly localized attention, while $6\times6\times6$ captures broader spatial information. All models were evaluated across different fine-tuning subject counts ($N$ = 1, 3, 6, 11).
\end{itemize}

All experiments were conducted on a server equipped with 8 NVIDIA V100 GPUs (32 GB memory each). Self-supervised pretraining was performed across all 8 GPUs using PyTorch 2.0.1 with CUDA 11.7, while fine-tuning was carried out on a single V100 GPU. For all training stages, we used the AdamW optimizer. The learning rate followed a consistent schedule: linear warmup during the first epoch, followed by cosine annealing.

\begin{table}[t]
\centering
\caption{Validation accuracies (Acc) of various augmentation strategies for pretraining on the HCP pretraining dataset. Please refer to \ref{data_aug_hcp} for explanation of \textit{High} and \textit{Low} settings. } 
\label{table:SSL_aug}
\setlength{\tabcolsep}{3pt}
\begin{tabular}{ccccc|c}
\toprule
\multicolumn{5}{c|}{\textbf{Augmentation strategies}} & \textbf{Acc (\%)} \\
\midrule
Noise & Smoothing & Striding & Masking & Affine & \\
\midrule
High & High & -- & -- & -- & 29.0 \\
High & High & \ding{51} & -- & -- & 54.9 \\
High & High & \ding{51} & \ding{51} & -- & 58.1 \\
High & High & \ding{51} & \ding{51} & \ding{51} & 50.0 \\
-- & -- & \ding{51} & -- & -- & 56.7 \\
-- & -- & \ding{51} & \ding{51} & -- & 58.6 \\
Low & Low & \ding{51} & \ding{51} & -- & \textbf{59.0} \\
\bottomrule
\end{tabular}
\end{table}

\begin{table}[t]
\centering
\caption{Augmentation strategies for downstream fine-tuning and their effects on accuracy and F1 scores. The reported average performance and standard deviation are obtained from experiments conducted on three different random splits.} 
\label{tab:mdtb_augmentation_results}
\setlength{\tabcolsep}{3pt}
\begin{tabular}{ccccc|cc}
\toprule
\multicolumn{5}{c|}{\textbf{Augmentation strategies}} & \textbf{Acc (\%)} & \textbf{F1 (\%)} \\
\midrule
Noise & Smoothing & Striding & Masking & Affine && \\
\midrule
Low & -- & -- & -- & -- & 58.7 (±3.42) & 57.6 (±3.66) \\
Low & Low & -- & -- & -- & 58.6 (±3.33) & 57.8 (±3.40) \\
Low & Low & \ding{51} & -- & -- & 58.8 (±2.81) & 57.8 (±2.82) \\
Low & Low & -- & \ding{51} & -- & 59.5 (±3.66) & 58.8 (±3.77) \\
Low & Low & -- & -- & \ding{51} & \textbf{62.4 (±5.28)} & \textit{61.4 (±6.03)} \\
Low & Low & -- & \ding{51} & \ding{51} & \textit{62.3 (±5.89)} & \textbf{61.6 (±6.44)} \\
\bottomrule
\end{tabular}

\end{table}

\begin{figure*}[t!]
    \centering
    \includegraphics[width=\textwidth]{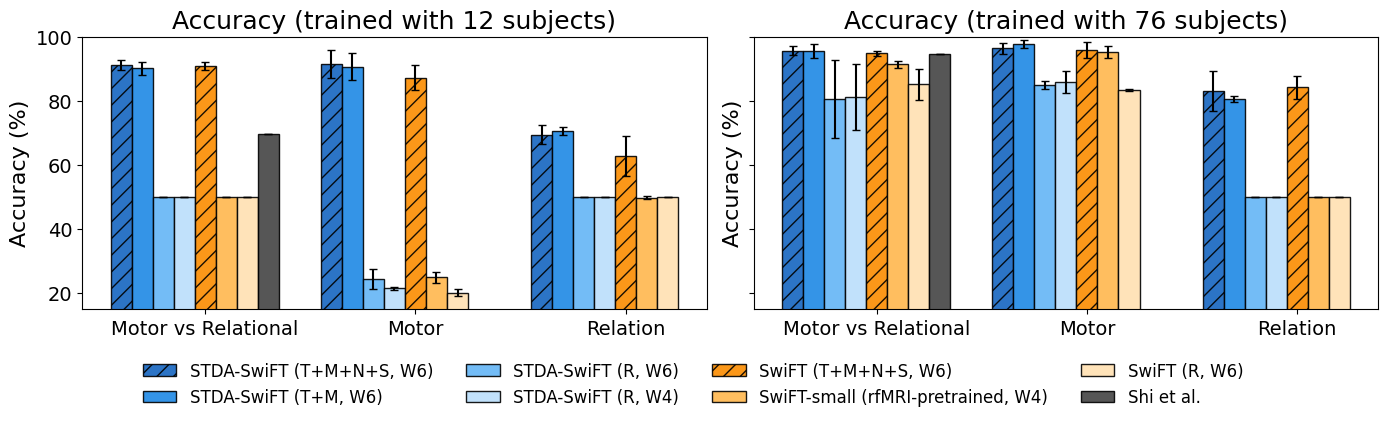}
    \caption{
        Fine-tuning performance on held-out HCP dataset. Models are annotated with their pretraining strategies and window sizes in parentheses. 
        \textbf{Pretraining strategies}: 
        \textbf{T} = Temporal striding, 
        \textbf{M} = Random masking, 
        \textbf{N} = Additive random noise, 
        \textbf{S} = Random smoothing, 
        \textbf{R} = Random initialization (no pretraining).  
        \textbf{Window sizes}: 
        \textbf{W4} = 4$\times$4$\times$4, 
        \textbf{W6} = 6$\times$6$\times$6.  
        For example, \textit{STDA-SwiFT (T+M+N+S, W6)} refers to our proposed STDA-SwiFT model trained with temporal striding, random masking, additive random noise, and random smoothing using a 6$\times$6$\times$6 window size. Additionally, SwiFT-small, released by \cite{kim2024swift} is pretrained on resting-state fMRI data. The original result of Shi \emph{et al.} \cite{shi2023self} is also reported.
    }
    \label{fig:hcp_fituning}
\end{figure*}

\section{Results and Discussion}

\subsection{Task 1}

Table~\ref{table:SSL_aug} shows the validation accuracy on the HCP validation set using various data augmentation strategies. We first note that unlike training image classifiers, simple augmentations such as noise and smoothing (row 1) yielded low accuracy (29.0\%), indicating limited effectiveness for task fMRI. Introducing random temporal striding (row 3) improved accuracy to 54.9\%, and using striding alone (row 4) achieved 56.7\%, highlighting its utility for self-supervised learning. Adding masking (row 6) further boosted performance to 58.1\%. However, applying affine transform (row 5) reduced accuracy to 50.0\%. The best result (59.0\%) was obtained by combining low-intensity noise/smoothing, striding, and masking (row 7). This setting was used in all subsequent experiments.

Table~\ref{tab:mdtb_augmentation_results} shows the fine-tuning accuracy and F1 scores on the MDTB test set after augmentation. Unlike pretraining, affine transform and masking yielded noticeable gains in fine-tuning. For example, adding affine transform to the noise+smooth setting improved accuracy from 58.6\% (row 2) to 62.4\% (row 6) and F1 from 57.8\% to 61.4\%. Notably, while temporal striding was most effective in pretraining (rows 3–4), it offered no improvement during fine-tuning. These results suggest that affine transform and masking were better in enhancing generalization in downstream tasks, whereas the benefit of striding was limited to pretraining.

\subsection{Task 2}

Figure~\ref{fig:hcp_fituning} shows the accuracy and F1 scores across various fine-tuning settings on the HCP held-out dataset. Models with pretraining consistently outperformed those without across all tasks. Among them, our proposed STDA-SwiFT model (leftmost two bars in each group) achieved the best performance overall. In the ``Motor vs. Relational'' task with only 12 training subjects (left), STDA-SwiFT achieved 91.0\% and 90.2\% accuracy, which substantially outperformed Shi \emph{et al.}~\cite{shi2023self} ($\sim$70\%). With 76 subjects (right), STDA-SwiFT (T+M, W6) further improved to 95.6\% and 95.7\%, surpassing the 94.5\% reported by Shi \emph{et al.} (trained with 200 subjects) and the fine-tuned SwiFT baseline (91.4\%). Similar trends were observed in the more challenging Motor and Relational tasks, where STDA-SwiFT maintained superior performance with lower computational cost. We also compared kernel sizes (third and fourth bars): performance with $6\times6\times6$ and $4\times4\times4$ windows was generally comparable except in the 12-subject Motor task, which remained most challenging.

These results demonstrate that our proposed STDA-SwiFT model consistently outperformed existing models with fewer training samples and less computational resources, and highlight the importance of appropriate data augmentation.

\begin{figure}[htbp]
    \centering
    \includegraphics[width=\linewidth]{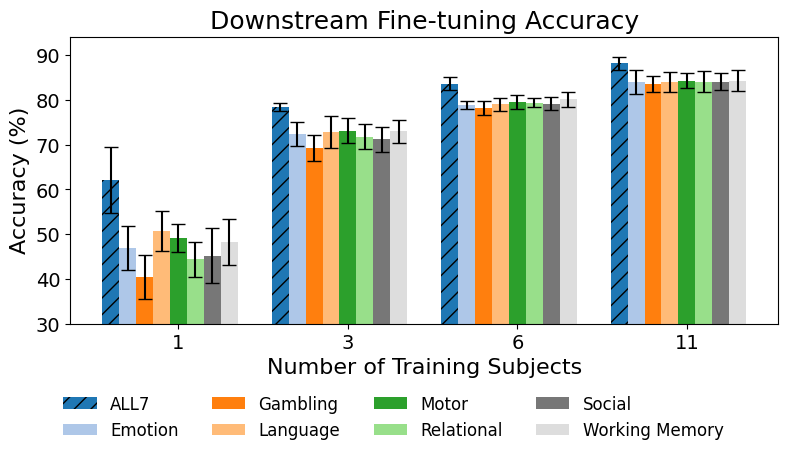}
    \caption{MDTB downstream fine-tuning performance for pretraining tasks with varying numbers of fine-tuning subjects.}
    \label{fig:MDTB_comparison}
\end{figure}

\subsection{Task 3}
Figure~\ref{fig:MDTB_comparison} shows the fine-tuning performance on the MDTB dataset using seven different HCP pretraining tasks (i.e., Emotion, Gambling, Language, Motor, Relational, Social, Working Memory, and ALL7) and subject counts ($N$ = 1, 3, 6, 11). As expected, the model pretrained on all seven tasks (ALL7) consistently achieved the highest accuracy (from 62.1\% at $N$ = 1 to 88.2\% at $N$ = 11). Among single-task pretraining, the language task performed best at $N$ = 1. However, as $N$ increased, performance across all single-task models converged ($\sim$83\%), although as expected multi-task pretraining (ALL7) retained a clear advantage.

\begin{table*}[ht]
\centering
\resizebox{\textwidth}{!}{
\begin{tabular}{l|c|cc|cc|cc|cc}
\toprule
\multirow{2}{*}{Framework} &\multirow{2}{*}{Method}& \multicolumn{2}{c|}{\textbf{N = 1}} & \multicolumn{2}{c|}{\textbf{N = 3}} & \multicolumn{2}{c|}{\textbf{N = 6}} & \multicolumn{2}{c}{\textbf{N = 11}} \\
\cline{3-10}
&& \textbf{Accuracy (\%)} & \textbf{F1 (\%)} & \textbf{Accuracy (\%)} & \textbf{F1 (\%)} & \textbf{Accuracy (\%)} & \textbf{F1 (\%)} & \textbf{Accuracy (\%)} & \textbf{F1 (\%)} \\
\midrule
Autoencoding \cite{thomas2022self} & Parcellation & 68.8 & 55.8 & 78.4 & 70.8 & 83.4 & 77.8 & 86.7 & 83.4 \\
CSM \cite{thomas2022self} & Parcellation & \textbf{77.1} & \textbf{69.9} & \textbf{85.1} & \textbf{83.1} & \textbf{88.5} & \textbf{87.1} & \textbf{90.0} & \textbf{89.7} \\
Net-BERT \cite{thomas2022self} & Parcellation & 71.6 & 54.7 & 81.1 & 72.1 & 84.6 & 78.8 & 87.5 & 84.3 \\
Seq-BERT \cite{thomas2022self} & Parcellation & 63.1 & 36.9 & 75.6 & 57.9 & 82.5 & 72.5 & 86.3 & 79.6 \\
\midrule
Shi \emph{et al.} \cite{shi2023self}\textsuperscript{†} & ROI & N/A & 31.0 & N/A & N/A & N/A & N/A & N/A & N/A \\
SwiFT \cite{kim2024swift} 4$\times$4$\times$4$\times$4 & Whole brain & 7.1 (±0.91) & 3.3 (±0.54) & 36.0 (±5.88) & 34.2 (±6.85) & 58.9 (±1.54) & 58.0 (±2.01) & 74.5 (±1.08) & 74.2 (±0.76) \\
\midrule
STDA-SwiFT 2$\times$2$\times$2 & Whole brain & 25.7 (±3.98) & 22.3 (±4.85) & 66.4 (±1.45) & 66.2 (±0.93) & 76.8 (±0.93) & 76.5 (±1.26) & 81.9 (±1.04) & 82.3 (±1.20) \\
STDA-SwiFT 4$\times$4$\times$4 & Whole brain & 11.9 (±1.26) & 7.7 (±1.18) & 63.6 (±1.70) & 63.4 (±2.16) & 77.2 (±0.69) & 77.1 (±0.98) & 82.0 (±1.60) & 82.0 (±1.87) \\
STDA-SwiFT 6$\times$6$\times$6 & Whole brain & 10.9 (±1.90) & 7.0 (±1.46) & 62.8 (±3.12) & 62.5 (±3.40) & 76.0 (±0.98) & 75.7 (±1.45) & 82.2 (±1.70) & 82.1 (±1.77) \\
STDA-SwiFT 6$\times$6$\times$6\textsuperscript{‡}  & Whole brain & \textbf{62.3} (±5.89) & \textbf{61.6} (±6.44) & \textbf{78.5} (±0.68) & \textbf{78.2} (±0.76) & \textbf{83.7} (±1.25) & \textbf{83.8} (±1.36) & \textbf{88.2} (±1.20) & \textbf{88.2} (±1.12) \\
\bottomrule
\end{tabular}
}

\caption{Fine-tuning performance on cognitive tasks in the MDTB dataset\cite{king2019functional}  with different training subject sizes and frameworks. The second and third blocks (Shi \emph{et al.}\ and SwiFT models) correspond to whole-brain models. \textsuperscript{†}In Shi \emph{et al.}~\cite{shi2023self}, `ROI' refers to model trained only on voxels in the visual cortex rather than whole-brain. \textsuperscript{‡} Model was pretrained on all 7 HCP tasks; all other STDA-Swift models were trained from scratch.}
\label{tab:mdtb_table}
\end{table*}

The upper part of Table~\ref{tab:mdtb_table} compares our STDA-SwiFT (ALL7, $6\times6\times6$) to ROI-/parcellation-based methods. Our method outperformed all except CSM~\cite{thomas2022self}, which required extensive processing to summarise activity in each parcel after atlas-based parcellation. By taking whole-brain data with minimal preprocessing, our model significantly reduced pipeline complexity whilst maintaining competitive performance. 

Examining performance across different fine-tuning dataset sizes, we also note that as $N$ increases, all methods improve, and the performance gap between our method and CSM narrows. We further showed that pretraining was critical: at $N=1$, the pretrained model greatly outperformed the randomly-initialized variant; at $N=11$, only the pretrained model outperformed most baselines, while the randomly initialized model did not.

The lower part of Table~\ref{tab:mdtb_table} also summarizes the effect of window size. STDA-SwiFT consistently outperformed SwiFT across all subject counts, which highlights the benefit of STDA over 4D-attention. We moreover find that our model intially performed better with smaller windows (e.g., $2\times2\times2$) when data was limited. At $N=1$, it achieved 25.7\% accuracy and 22.3\% F1, compared to 10.9\% and 7.0\% for the $6\times6\times6$ variant.

However, this gap narrowed as $N$ increased. At $N=11$, STDA-Swift with $6\times6\times6$ window size reached 82.2\% accuracy and 82.1\% F1, which was comparable results trained with smaller-window sizes.

These results suggest that smaller windows introduce useful local inductive bias for low-data regimes, while larger windows benefit from richer data to model global patterns more effectively.

\begin{figure*}[htbp]
    \centering
    \includegraphics[width=0.9\linewidth]{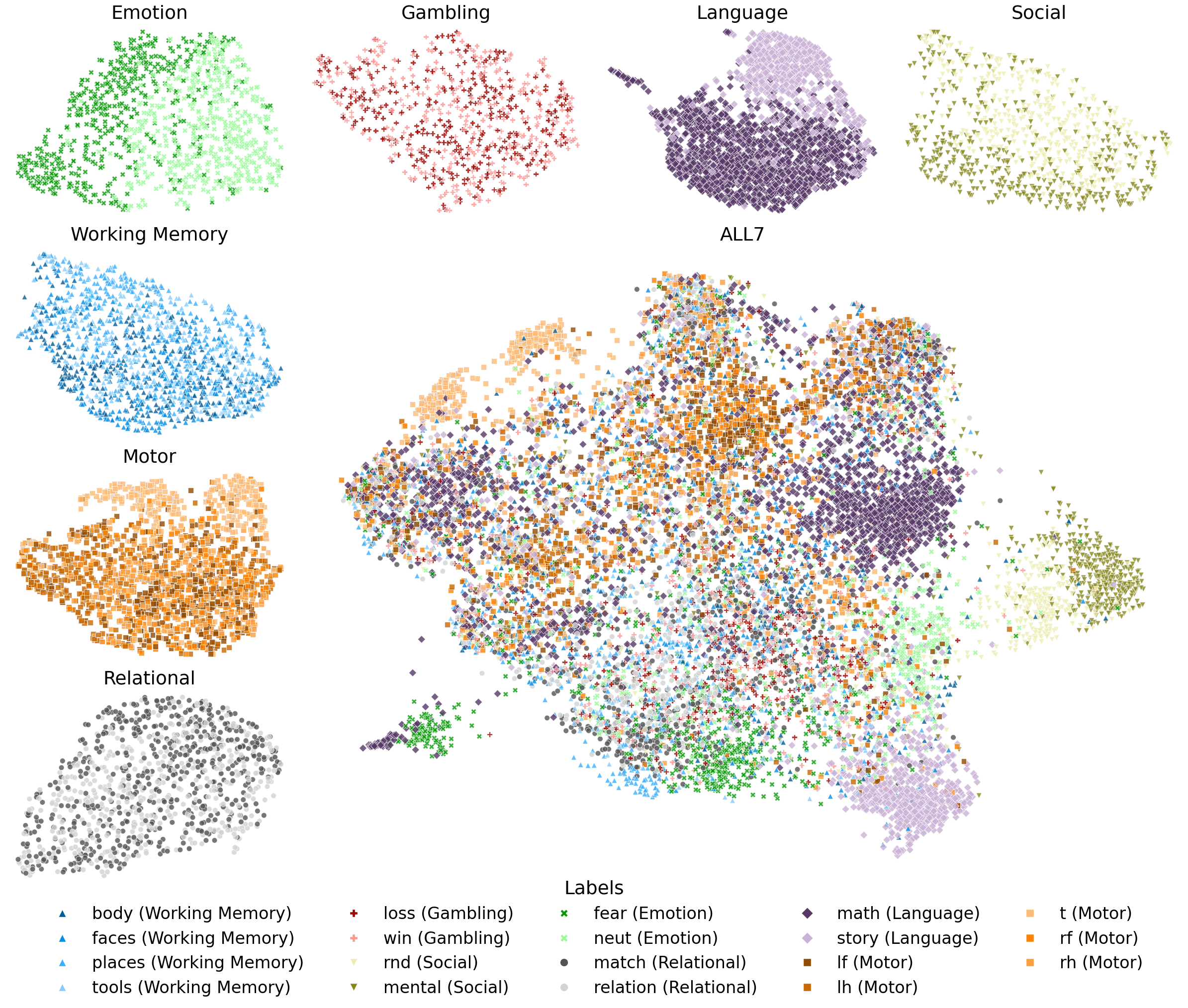}
    \caption{UMAP of learned condition embeddings for various tasks.}
    \label{fig:umap_embeddings}
\end{figure*}

\subsection{Features learned from self-supervised learning}
Figure~\ref{fig:umap_embeddings} shows the feature embeddings produced by our STDA-SwiFT model with a $6\times6\times6$ window size, pretrained on the seven tasks from the HCP dataset. The embeddings are visualized using Uniform Manifold Approximation and Projection (UMAP) \cite{mcinnes2020umap}, a dimensionality reduction technique that enables visualization of high-dimensional data in a lower-dimensional space. Each point represents a mental state from one subject, and colors correspond to different task conditions. This visualization allows us to qualitatively assess how the model organizes brain activity patterns, revealing task-specific structures and relationships in the learned feature space.

Examining models pretrained on individual tasks, we found that certain task pairs (e.g., math/story, fear/neutral, rnd/mental, and t/others) formed more separable embeddings than others (e.g. match/relation, loss/win). When pretrained on all seven tasks (ALL7), these distinctions largely persisted with additional emergent clusters, such as math/fear in the lower-left corner. Classes in the gambling, working memory, and relational task remained, likely because they engage multiple cognitive processes (e.g., decision-making, memory retrieval, abstract reasoning) and the variable nature of their neural representations, which make a clear separation in the learned feature space challenging.

\begin{figure}[htbp]
    \centering
    \includegraphics[width=\linewidth, trim={0 0 0 1cm}, clip]{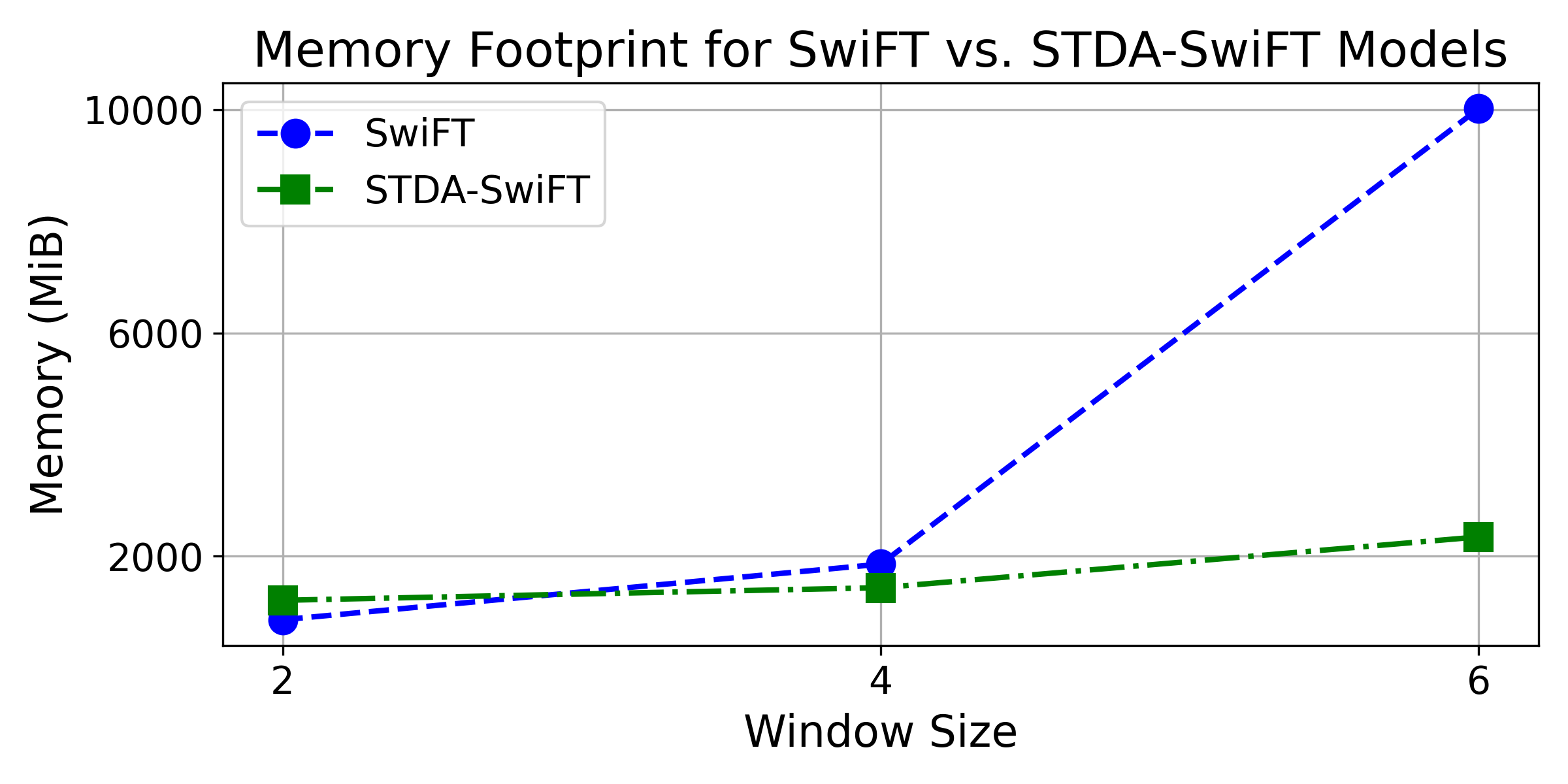}
    \caption{Comparison of memory footprint for the SwiFT and STDA-SwiFT models across different window sizes.}
    \label{fig:memory_comparison}
\end{figure}

\subsection{Memory Footprint}  

Finally, we compared the memory footprint of our proposed STDA-SwiFT model under various spatial window sizes against SwiFT as a baseline to assess the efficiency of the STDA mechanism. Both models were evaluated using a standardized input of shape \( (1, 1, 96, 96, 96, 15) \) in \texttt{float16} precision, simulating a single fMRI volume with 15 time frames. Memory usage during a single forward pass was recorded while varying spatial window sizes (2$\times$2$\times$2, 4$\times$4$\times$4, 6$\times$6$\times$6), with temporal window fixed at 15, on an NVIDIA V100 GPU.

Figure~\ref{fig:memory_comparison} summarizes the results. At smaller window sizes (e.g., \(2 \times 2 \times 2\)), the 4D model consumes 859~MiB, while the STDA model (with a split-window configuration of \(2 \times 2 \times 2 \times 1 + 1 \times 1 \times 1 \times 15\)) requires 1,207~MiB. However, as window size increases, the 4D model's memory usage scales steeply, reaching 10,021~MiB at \(6 \times 6 \times 6\), while STDA remains significantly lower at 2,343~MiB. This confirms the scalability and memory efficiency of the STDA-SwiFT model for whole-brain decoding.

\section{Conclusion}


In this study, we present \textbf{STDA-SwiFT}, an efficient model for self-supervised representation learning on whole-brain task fMRI data. Compared to prior work, our approach introduces three main advances: First, the \textbf{spatio-temporal decoupled attention design} significantly reduces memory consumption compared to fully 4D attention models such as SwiFT, and we further show that smaller spatial window sizes are particularly beneficial when the amount of fine-tuning data is limited. Second, we propose a \textbf{simple yet effective contrastive learning strategy} for fMRI pretraining, which leverages random temporal striding and masking without relying on handcrafted region-of-interest (ROI) selection or parcellation-based feature extraction. Inspired by the spatiotemporal nature in video understanding \cite{wang2022long}, this design makes our pipeline fully end-to-end and broadly applicable to whole-brain inputs. Third, our pretrained STDA-SwiFT model demonstrates strong \textbf{transferability across datasets}, achieving competitive performance on the MDTB dataset despite domain shifts, highlighting the robustness of both the model and the learned representations. Together, our findings suggest that combining architectural efficiency with tailored contrastive strategies can enable practical, scalable, and effective whole-brain fMRI decoding. 

\section*{Acknowledgment}
We thank the National Center for High-performance Computing (NCHC) of National Applied Research Laboratories (NARLabs) in Taiwan for providing computational and storage resources.


\bibliographystyle{IEEEtran}
\bibliography{references}

\end{document}